\documentclass[11pt, oneside,onecolumn]{article}   	
\usepackage{geometry}                		
\usepackage{float}
\usepackage[caption = false]{subfig}
\usepackage{graphicx}				
							
\usepackage{amssymb, amsmath}
\usepackage{enumerate}

\newtheorem{thm}{Theorem}[section]

\newtheorem{dfn}{Definition}[section]

\newtheorem{prf}{Proof}[section]

\title{Orthogonal Ramanujan Sums, its properties and Applications in Multiresolution Analysis}

\author{ Devendra Kumar Yadav, Gajraj Kuldeep, S. D. Joshi  }

\date{}							

\begin{document}

\maketitle

\title{\textbf {Abstract:-}} Signal processing community has recently shown interest in Ramanujan sums which was defined by S.Ramanujan in 1918. In this paper we have proposed Orthogonal Ramanujan Sums (ORS) based on Ramanujan sums. In this paper we present two novel application of ORS. Firstly a new representation of a finite length signal is given using ORS which is defined as Orthogonal Ramanujan Periodic Transform.Secondly ORS has been applied to multiresolution analysis and it is shown that Haar transform is a special case.\\

\section{Introduction}

	The great Indian mathematician S.Ramanujan defined a trigonometric sum ~\cite{bib:9}  as
\begin{equation}
	c_q(n) = \sum_{\begin{subarray}{c}
	                         k=1\\
	                         (k, q) =1\\
	                         \end{subarray}}^{q}\exp^{\frac{j2\pi kn}{q}}
\end{equation}

where $(k,q)=1$ implies that k and q are relatively prime. 
Various standard arithmetic functions like Mobius function $\mu(n)$, Euler's toient function $\phi(n)$,von Mangoldt function , Riemann-zeta function $\zeta(s)$ were represented using these Ramanujan sums. Further Ramanujan proved that any signal $x(n)$ can be represented by a linear combination of $c_q(n)$ which is called as Ramanujan Sum Expansion(RSE).

\begin{equation}
	x(n) = \sum\limits_{q=1}^{\infty} {\alpha} _q c_q(n)
\end{equation}
Where $\alpha_q's$ are RSE coefficients.Since the limits in the equation(2) are infinite therefore number of $c_q(n)$ required to represent any arbitrary function $x(n)$ would be infinite. Carmichael ~\cite{bib:8} gave the formula below for the calculation of the RSE coefficients $\alpha_q's$

\begin{equation}
	\alpha_q = \frac{1}{\phi(q)}\left(\lim_{M \rightarrow \infty} \sum\limits_{n=1}^{M} {x(n) c_q(n)} \right)
\end{equation}
M.Planat et al. showed that these sums can be used for analysis of long-period sequences and $\frac{1}{f}$ noise ~\cite{bib:4}~\cite{bib:5}. M.Planat et al. in ~\cite{bib:5} have shown that the result obtained in ~\cite{bib:4} was quite ambiguous.Authors ~\cite{bib:1} have recently shown that Ramanujan Sums behave as first derivative which gives more insight to Ramanujan Sums.
P.P. Vaidyanathan ~\cite{bib:6} ~\cite{bib:7} gave an impetus by giving two representations Ramanujan FIR Transform and Ramanujan Periodic Transform based on Ramanujan sums. Further he gave an application of these transforms to find the hidden period of a finite length signal. In the first representation any arbitrary signal of length N is represented by linear combination of first N Ramanujan sums. He further showed that this approach for determining period is sensitive to the shift in the input signal. In the second representation he defined Ramanujan Periodic Transform(RPT) using Ramanujan subspaces. This transform is able to find hidden periods of the given signal of length N. But it also has a limitation that it is not able to find the hidden period other than the divisors of N. Both RFT and RPT have been used to find the periodic structure in the signal but has not been explored in other signal processing applications.\\
Recently authors have shown application of Ramanujan sums in image processing in particular for finding edges and noise level estimation\cite{bib:1}. In this paper a new property of Ramanujan sums has been derived. This property makes it suitable for multirate signal processing. It shows that higher order Ramanujan Sums are interpolated version of lower order Ramanujan sums. In this paper,we have defined Orthogonal Ramanujan Sums (ORS) based on Ramanujan sums. Higher order ORS are also interpolated version of lower orders by its definition . These are very useful in signal processing applications. Two novel application of ORS have been presented in this paper. A new representation of a finite length signal is given using ORS which is defined as Orthogonal Ramanujan Periodic Transform.Application of ORS in generalising Discrete Wavelet Transform is demonstrated in this paper. Generally MRA is applied on dyadic scale, here instead of representing it on dyadic scale, we propose that it can be applied at any scale. For e.g. suppose $N=PM$ is the length of the signal and $P >  1$, we can generate $P$-scale MRA of the given signal where one component will give the average of the signal and $P-1$ components will give the detail part of the signal.When $P=2$, we get Haar transform, as a particular case.\\
The organisation of this paper is as follows: In section 2 Orthogonal Ramanujan Sums(ORS)  are defined and some of its properties are proved. Orthogonal Ramanujan Periodic Transform(ORPT) is defined in section 3. Section 4 contains Generalised Discrete Wavelet Transform using ORPT. Results are illustrated in sections followed by concluding remarks.

\section{Orthogonal Ramanujan Sums (ORS) }

In this section we propose a family of sequences, termed here as Orthogonal Ramanujan Sums (ORS), and discuss some of its properties.\\
\begin{dfn}
Let us denote, for any prime $q$,\\ $c^{k}_q(n) = u(((n))_q-k)c_q(n-k)-k\delta(((n))_q-k)$, where $((n))_q$ is $n$ mod $q$,\\ $ u(n-k) $ =$\left\{\begin{array}{cc}
											1 & n \ge k\\
											0 & elsewhere
											\end{array}\right.$ and $\delta(n-k)$ =$\left\{\begin{array}{cc}
											1 & n = k\\
											0 & elsewhere
											\end{array}\right.$\\  
for $0 \le k < q-1$.
\end{dfn}
For example\\

\begin{tabular}{lcl} $\textbf{q = 3}$ & &$\textbf{q = 5}$ \\ $ c^0_3(n) = [2,-1,-1]$ & & $c^0_5(n) = [4,-1,-1,-1,-1]$\\ $c^1_3(n) = [0,1,-1]$ & & $c^1_5(n) = [0,3,-1,-1,-1]$\\ & & $c^2_5(n)=[0,0,2,-1,-1]$\\ & & $c^3_5(n)=[0,0,0,1,-1]$\end{tabular}\\


It can be observed that $c_q^k(n)$ is periodic in $n$ with period $q$. Also $c_q^0(n)$ is same as Ramanujan sum for a given prime $q$. Further it is clear that terms in $c_q^k(n)$ are integer only. Next we will prove that sum of $c_q^k(n)$ is zero over a period and its energy is finite.
\begin{thm}
$\sum_{n=0}^{q-1}c^{k}_q(n) = 0$ and $\sum_{n=0}^{q-1}(c^{k}_q(n))^2 = (q-k)(q-k-1)$ $ \forall $ $0 \le k < q-1$
\end{thm}

\begin{prf}
From definition
\begin{eqnarray*}
	\sum_{n=0}^{q-1}c^{k}_q(n) = \sum_{n=0}^{q-1}(u(((n))_q-k)c_q(n-k)-k \delta(((n))_q-k))\\
	=\sum_{n=k}^{q-1}c_q(n-k)-k = 0
\end{eqnarray*}	
($\because$ sum of $c_q(n)$ is zero over one period and for prime q we know that $c_q(n) = -1$ when $n \ne 0$ for one period)
This completes the first part.\\
For prime q,$c^{k}_q(n)$ has a particular form of $k$ zeroes followed by $q-1-k$ and rest of the entries as $-1$. So the norm is equal to $(q-1-k)^2+(q-1-k)$
This proves theorem $2.1.$
\end{prf}
As we know that Ramanujan sum $c_{q_1}(n)$ and $c_{q_2}(n)$ are orthogonal over lcm($q_1,q_2$). Here we will prove that  $c^{k}_q(n)$ are orthogonal for a fixed q and different values of $k$.

\begin{thm}
For a prime q, $c^{k}_q(n)$'s are orthogonal for $0 \le k < q-1$.
\end{thm}

\begin{prf}
For   $0 \le n < q-1$
\begin{eqnarray*}
	c^{k_1}_q(n) = u(n-k_1)c_q(n-k_1)-k_1\delta(n-k_1)\\
	c^{k_2}_q(n) = u(n-k_2)c_q(n-k_2)-k_2\delta(n-k_2),k_1 \ne k_2
\end{eqnarray*}
Taking inner product, we get

\begin{eqnarray*}
	<c^{k_1}_q(n) , c^{k_2}_q(n)>  = \sum_{n=0}^{q-1} c^{k_1}_q(n),c^{k_2}_q(n)
\end{eqnarray*}

\begin{eqnarray*}
	<c^{k_1}_q(n) , c^{k_2}_q(n)> = \sum_{n=0}^{q-1} u(n-k_1)u(n-k_2)c_q(n-k_1) c_q(n-k_2)
						    -\sum_{n=0}^{q-1} k_1 u(n-k_2) c_q(n-k_2)\delta(n-k_1)\\
						    -\sum_{n=0}^{q-1} k_2 u(n-k_1) c_q(n-k_1)\delta(n-k_2)
						    +\sum_{n=0}^{q-1}k_1k_2\delta(n-k_1)\delta(n-k_2)
\end{eqnarray*}

	Assuming $k_1 > k_2$,we get

\begin{eqnarray*}
	<c^{k_1}_q(n) , c^{k_2}_q(n)> = \sum_{k_1}^{q-1} c_q(n-k_1) c_q(n-k_2) - k_1c_q(k_1-k_2)
\end{eqnarray*}

	Since $k_1 > k_2$ and $q$ is prime,therefore $c_q(k_1-k_2)=-1$.Therefore

\begin{eqnarray*}
	<c^{k_1}_q(n) , c^{k_2}_q(n)> = \sum_{n=0}^{k_1-1}c_q(n-k_1) + k_1 =0
\end{eqnarray*}

\end{prf}

\textbf{Note}: Since $c^{k}_q(n)$ have been derived from Ramanujan sums and are orthogonal hence these are termed here as Orthogonal Ramanujan Sequences(ORS).\\
Since ORS are defined only for prime, next we will show how these can be used to generate ORS for composites.
\begin{dfn}
For $q_1$ and $q_2$ to be two distinct prime, we can define a new orthogonal Ramanujan sequences $c^{k_1 k_2}_{q_1 q_2}(n) = c^{k_1}_{q_1}(n)c^{k_2}_{q_2}(n) $, where $k_1$ and $k_2$  are shifts of $c^{k_1}_{q_1}(n)$ and $c^{k_2}_{q_2}(n)$ respectively. 
\end{dfn}

\begin{thm}
Let $c^{k_1 k_2}_{q_1 q_2}(n)$ be as defined above. Then
\begin{enumerate}
\item $<c^{k_1}_{q_1}(n) , c^{k_2}_{q_2}(n)>  = 0$
\item $<c^{k_1 k_2}_{q_1 q_2}(n) , c^{k_1}_{q_1}(n)> = 0$ 
\item $<c^{k_1 k_2}_{q_1 q_2}(n) , c^{k_2}_{q_2}(n)> = 0$
\end{enumerate}
for $0 \le n \le q_1q_2-1$.
\end{thm}

\begin{prf}
From definition $c^{k_1}_{q_1}(n)$ and $c^{k_2}_{q_2}(n)$ are periodic with periods $q_1$ and $q_2$  respectively. 
\begin{eqnarray*}
<c^{k_1}_{q_1}(n) , c^{k_2}_{q_2}(n)> = \sum_{n=0}^{q_1q_2-1} c^{k_1}_{q_1}(n)c^{k_2}_{q_2}(n)
\end{eqnarray*}
This can be further reduced to 
\begin{eqnarray*}
=\sum_{n=0}^{q_1-1} c^{k_1}_{q_1}(n) \sum_{n=0}^{q_2-1}c^{k_2}_{q_2}(n)
\end{eqnarray*}
As we know that
\begin{eqnarray*}
\sum_{n=0}^{q-1} c^{k}_{q}(n)=0
\end{eqnarray*}
This proves the claim that $c^{k_1}_{q_1}(n)$ and $c^{k_2}_{q_2}(n)$ are orthogonal.\\
Similarly it can be shown that $c^{k_1 k_2}_{q_1 q_2}(n)$ is orthogonal to $c^{k_1}_{q_1}(n)$ and $c^{k_2}_{q_2}(n) $.
\end{prf}

An important property of Ramanujan sequence is derived here.

\begin{thm}
For a prime $q$,
	$c_{q^l}(n) = q^{l-1}c_{q}(\frac{n}{q^{l-1}})$ ,where $l > 0$
\end{thm}

\begin{prf}
From definition of $c_q(n)$, $c_{q^l}(n)$ can be written as \\
\begin{eqnarray*}
	c_{q^l}(n) = \sum_{\begin{subarray}{c}
	                         k=1\\
	                         (k, q) =1\\
	                         \end{subarray}}^{q^l}\exp^{\frac{j2\pi kn}{q^l}}
\end{eqnarray*}

Let $r_1,r_2,\hdots,r_s$ be the terms which are relatively prime to $q$. Therefore other terms which are relatively prime to $q^l$ are of the form\\
\begin{math}
\begin{array}{cccc}
q+r_1, & q+r_2, & \hdots & ,q+r_s \\
2q+r_1, & 2q+r_2, & \hdots & ,2q+r_s \\
\vdots &\vdots & \vdots &\vdots \\
(q^{l-1}-1)q+r_1, &(q^{l-1}-1)q+r_2, &\hdots & ,(q^{l-1}-1)q+r_s \end{array}\\
\end{math}
Using this, rewriting the above expression of $c_{q^l}(n)$

\begin{eqnarray*}
c_{q^l}(n) = \exp^{\frac{j2\pi r_1n}{q^l}}(1+\exp^{\frac{j2\pi n}{q^{l-1}}}+\hdots+\exp^{\frac{j2\pi (q^{l-1}-1)n}{q^{l-1}}}) \hspace{0.5in} \\
	+\exp^{\frac{j2\pi r_2n}{q^l}}(1+\exp^{\frac{j2\pi n}{q^{l-1}}}+\hdots +\exp^{\frac{j2\pi (q^{l-1}-1)n}{q^{l-1}}}) \hspace{0.5in} \\
  \vdots 	\hspace{1.5in}\\
	+\exp^{\frac{j2\pi r_sn}{q^l}}(1+\exp^{\frac{j2\pi n}{q^{l-1}}}+\hdots +\exp^{\frac{j2\pi (q^{l-1}-1)n}{q^{l-1}}}) \hspace{0.5in}\\
\end{eqnarray*}	

Since
$(1+\exp^{\frac{j2\pi n}{q^{l-1}}}+\exp^{\frac{j2\pi 2n}{q^{l-1}}}+\hdots +\exp^{\frac{j2\pi (q^{l-1}-1)n}{q^{l-1}}})$ = $\left\{\begin{array}{cc} q^{l-1} & \text{when } q^{l-1}|n \\ 0 & \text{else} \end{array}\right.$


Therefore
\begin{eqnarray*}
	c_{q^l}(n) = (\exp^{\frac{j2\pi r_1n}{q^l}}+\exp^{\frac{j2\pi r_2n}{q^l}}+\hdots+\exp^{\frac{j2\pi r_sn}{q^l}})q^{l-1}\\
	=q^{l-1}\sum_{\begin{subarray}{c}
	                         k=1\\
	                         (k, q) =1\\
	                         \end{subarray}}^{q}\exp^{\frac{j2\pi kn}{q^l}}\\
	= q^{l-1}c_{q}(\frac {n}{q^{l-1}})
\end{eqnarray*}	

\end{prf}

This proves that higher order Ramanujan sums are interpolated versions of lower order Ramanujan sums. Since any integer $N$ can be written as $p_1^{r_1}p_2^{r_2} \hdots p_m^{r_m}$. Thereby using multiplicative property of Ramanujan sequences we can write

\begin{eqnarray*}
	c_N(n) = c_{p_1^{r_1}}(n)c_{p_2^{r_2}}(n) \hdots c_{p_m^{r_m}}(n)
\end{eqnarray*}

\textbf{Corollary:} Using above theorem $c_N(n)$ can be written as
\begin{eqnarray*}
	c_N(n) = p_1^{r_1-1}p_2^{r_2-1} \hdots p_m^{r_m-1} c_{p_1}(\frac {n}{p_1^{r_1-1}})c_{p_2}(\frac {n}{p_2^{r_2-1}}) \hdots c_{p_m}(\frac {n}{p_m^{r_m-1}})
\end{eqnarray*}
%
%
%

Thus any Ramanujan sequence, $c_N(n)$, can be represented as an interpolation of Ramanujan sequences for prime divisors of $N$ ~\cite{bib:10}. Similarly interpolation of Orthogonal Ramanujan Sequences can be represented in same way.i.e,
\begin{dfn}

For a prime $q$,
	$c^k_{q^l}(n) = q^{l-1}c^k_{q}(\frac{n}{q^{l-1}})$, where $l > 0$ and $0 \le k < q -1$.
\end{dfn}
From definition it is clear that interpolated ORS $c^k_{q^l}(n)$ are also orthogonal for different values of $k$.
\begin{thm}
For arbitrary positive integer N, Orthogonal Ramanujan Sequences can be expressed in terms of Orthogonal Ramanujan Sequences of its prime factors as
\begin{eqnarray*}
c^{k_1k_2 \hdots k_m}_N(n) = p_1^{r_1-1}p_2^{r_2-1} \hdots p_m^{r_m-1} c^{k_1}_{p_1}(\frac {n}{p_1^{r_1-1}})c^{k_2}_{p_2}(\frac {n}{p_2^{r_2-1}}) \hdots c^{k_m}_{p_m}(\frac {n}{p_m^{r_m-1}})
\end{eqnarray*}
\end{thm}

\begin{prf}
Similar to proof of theorem 2.3 above.
\end{prf}

Now we will prove that for any $N$ there are $\phi$(N) Orthogonal Ramanujan Sequences where $\phi$(N) is number of relatively prime numbers to $N$.  

\begin{thm}
For any $N>0$,there exists $\phi$(N) Orthogonal Ramanujan Sequences.
\end{thm}

\begin{prf}
For a particular $N = p_1^{r_1}$.By definition 
\begin{eqnarray*}
	c^{k_1}_{p_1^{r_1}}(n) = p_1^{r_1-1}c^{k_1}_{p_1}(\frac{n}{p_1^{{r_1}-1}})
\end{eqnarray*}


As we know that if we interpolate a signal $c^{k_1}_{p_1}(n)$by a factor of ${p_1}^{r_1-1}$ then $c^{k_1}_{{p_1}^{r_1}}(n-l)$ are pairwise orthogonal for $0 \le l \le {p_1}^{r_1-1}-1$ shifts. So total number of orthogonal sequences are ${p_1}^{r_1-1}$ for particular $k_1$. From theorem 2.2 we have $p_1-1$ orthogonal vectors from $c^{k_1}_{p_1}(n)$. Therefore total number of orthogonal vectors for all possible values of $k_1$ are ${p_1}^{r_1-1}({p_1-1})$ which is equal to $\phi ({p_1}^{r_1})$.
\\\\
Now for generalised $N$  = $p_1^{r_1}p_2^{r_2} \hdots p_m^{r_m}$, using above theorems, we can choose orthogonal  $c^{k_1k_2 \hdots k_m}_N(n)$ by $\phi ({p_1}^{r_1}) \phi ({p_2}^{r_2}) \hdots \phi ({p_m}^{r_m})$ which is equivalent to $\phi (N)$. This concludes the proof of theorem $2.6$.

\end{prf}

\section{Orthogonal Ramanujan Periodic Transform (ORPT)}
Using Orthogonal Ramanujan Sequences, Orthogonal Ramanujan Periodic Transform(ORPT) has been defined in this section.

\begin{thm}
Any arbitrary signal $x(n)$, of length $N$ can be represented as,
\begin{eqnarray*}
x(n)  = \sum_{d_i | N} \sum_{j_1=0}^{(p_{i1}^{r_{i1}-1}-1)} \sum_{j_2=0}^{(p_{i2}^{r_{i2}-1}-1)} \hdots \sum_{j_m=0}^{(p_{im}^{r_{im}-1}-1)} \sum_{k_1=0}^{\phi(p_{i1})-1} \sum_{k_2=0}^{\phi(p_{i2})-1} \hdots\sum_{k_m=0}^{\phi(p_{im})-1} (\beta_{d_i;j_1,j_2 \hdots ,j_m;k_1,k2 \hdots ,k_m} \\  {c^{k_1}_{p_{i1}}(\frac{n}{p_{i1}^{r_{i1}-1}}-j_1)}  {c^{k_2}_{p_{i2}}(\frac{n}{p_{i2}^{r_{i2}-1}}-j_2)} \hdots  {c^{k_m}_{p_{im}}(\frac{n}{p_{im}^{r_{im}-1}}-j_m)}
\end{eqnarray*}
where $d_i$'s are divisor of $N$ and each $d_i$ is of the form $p_{i1}^{r_{i1}}p_{i2}^{r_{i2}} \hdots p_{im}^{r_{im}}.$
\end{thm}

\begin{prf}
In the previous section we have seen that Orthogonal Ramanujan Sequences are pairwise orthogonal and contains integer entries. Also for a given $d$ we have $\phi (d)$ Orthogonal Ramanujan Sequences. As we know that ~\cite{bib:2} for any N;
\begin{eqnarray*}
N = \sum_{d_i | N}\phi (d_i)
\end{eqnarray*}
Where $d_i$ are divisors of $N$ .
Hence for any $N$ we have $N$ orthogonal sequences. Using these orthogonal sequences as a basis, any finite length signal $x(n)$ of length $N$ can be represented as shown in the theorem. This completes the proof.
\end{prf}
$\beta_{d_i;j_1,j_2 \hdots ,j_m;k_1,k_2 \hdots ,k_m} $ are  ORPT coefficients of $x(n)$ which can be represented as : \\

$\beta_{d_i;j_1,j_2 \hdots ,j_m;k_1,k_2 \hdots ,k_m}  = <x(n),({c^{k_1}_{p_{i1}}(\frac{n}{p_{i1}^{r_{i1}-1}}-j_1)}  {c^{k_2}_{p_{i2}}(\frac{n}{p_{i2}^{r_{i2}-1}}-j_2)} \hdots  {c^{k_m}_{p_{im}}(\frac{n}{p_{im}^{r_{im}-1}}-j_m)})>$

\begin{flushleft}
	\textbf{Example}
\end{flushleft}
Consider $N = 12$. Its divisors are $1,2,3,4,6,12$. Therefore $x(n)$ can be represented as

\begin{eqnarray*}
x(n)  = \beta_{1;0;0}c_{1}^{0}(n)+\beta_{2;0;0}c_{2}^{0}(n)+\beta_{3;0;0}c_{3}^{0}(n)+\beta_{3;0;1}c_{3}^{1}(n)+\beta_{4;0;0}c_{2}^{0}(\frac{n}{2})+\beta_{4;1;0}c_{2}^{0}(\frac{n}{2}-1) \\ +\beta_{6;0,0;0,0}c_{2}^{0}(n)c_{3}^{0}(n)+\beta_{6;0,0;0,1}c_{2}^{0}(n)c_{3}^{1}(n) +\beta_{12;0,0;0,0}c_{2}^{0}(\frac{n}{2})c_{3}^{0}(n)+\beta_{12;1,0;0,0}c_{2}^{0}(\frac{n}{2}-1)c_{3}^{0}(n) \\ +\beta_{12;0,0;0,1}c_{2}^{0}(\frac{n}{2})c_{3}^{1}(n)+\beta_{12;1,0;0,1}c_{2}^{0}(\frac{n}{2}-1)c_{3}^{1}(n)
\end{eqnarray*}
Orthogonal Ramanujan sequences can be normalised using theorem $2.1$. Similarly Orthonormal Ramanujan Periodic Transform of a finite length signal can be defined, in the normalised form as:
\begin{eqnarray*}
x(n)  = \sum_{d_i | N} \sum_{j_1=0}^{(p_{i1}^{r_{i1}-1}-1)} \sum_{j_2=0}^{(p_{i2}^{r_{i2}-1}-1)} \hdots \sum_{j_m=0}^{(p_{im}^{r_{im}-1}-1)} \sum_{k_1=0}^{\phi(p_{i1})-1} \sum_{k_2=0}^{\phi(p_{i2})-1} \hdots\sum_{k_m=0}^{\phi(p_{im})-1} (\beta_{d_i;j_1,j_2 \hdots ,j_m;k_1,k2 \hdots ,k_m} \\ \frac {c^{k_1}_{p_{i1}}(\frac{n}{p_{i1}^{r_{i1}-1}}-j_1)}{\sqrt((\frac{N}{p_{i1}^{r_{i1}}})(p_{i1}-k_1)(p_{i1}-k_1-1))} \frac {c^{k_2}_{p_{i2}}(\frac{n}{p_{i2}^{r_{i2}-1}}-j_2)}{\sqrt((\frac{N}{p_{i2}^{r_{i2}}})(p_{i2}-k_2)(p_{i2}-k_2-1))} \hdots \\ \frac {c^{k_m}_{p_{im}}(\frac{n}{p_{im}^{r_{im}-1}}-j_m)}{\sqrt((\frac{N}{p_{im}^{r_{im}}})(p_{im}-k_m)(p_{im}-k_m-1))})
\end{eqnarray*}
Since ORPT coefficients are inner product of signal and orthogonal Ramanujan sequences. Therefore signal $x(n)$ of length $N$ can be represented as 
\begin{eqnarray*}
x_N =  R_N \beta,  \hspace{1 cm}  \beta_N =[\beta_1,\beta_2, \hdots ,\beta_N]'
\end{eqnarray*}
where $x_N$ and $\beta_N$ are column vectors of size $N$x$1$ and $R_N$ be the $N$x$N$ matrix which is the equivalent representation of ORPT. By construction each column of $R_N$ are pairwise orthogonal.Hence $R_N$ is invertible and $\beta_N = R_N^{-1}x_N$. In an earlier work authors have shown that Ramanujan sequences are basically first order derivative ~\cite{bib:1}. Hence $\beta_1$ is smoothing coefficient and $\beta_N (N>1)$ can be interpreted as finer details of a signal. Few examples of $R_N$ matrix are \\

\begin{eqnarray*}
R_{2} = \left[\begin{array}{rr}1 &1\\
					    1 & -1\end{array}\right]
\end{eqnarray*}

\begin{eqnarray*}
R_{3} = \left[\begin{array}{rrr}1 & 2 & 0\\
					    1 & -1 & 1\\
					    1 & -1 & -1\end{array}\right]
\end{eqnarray*}

\begin{eqnarray*}
R_{4} = \left[\begin{array}{rrrr}1 & 1 & 1 &0\\
					    1 & -1 & 0 & 1\\
  					    1 &  1 & -1 & 0\\
					    1 &  -1 & 0 & -1\end{array}\right]
\end{eqnarray*}

\begin{eqnarray*}
R_{6} = \left[\begin{array}{rrrrrr}1 & 1 & 2 & 0 & 2 &0\\
					    1 & -1 & -1 & 1 & 1 &-1\\
					    1 & 1 & -1 & -1 & -1 &-1\\
					    1 & -1 & 2 & 0 & -2 &0\\
					    1 & 1 & -1 & 1 & -1 &1\\
					    1 &  -1 & -1 & -1 & 1 &1\end{array}\right]
\end{eqnarray*}
Observe that ORPT can also be used to find hidden periods of a finite length discrete signal,as was done in ~\cite{bib:6} because of the properties of the orthogonal Ramanujan sequences. 

\section{Generalised Discrete Wavelet transform}
In this section we will demonstrate that ORPT can be used to generalise Discrete Wavelet Transform (DWT) in finite dimension. In DWT a signal is decomposed into two parts: (i) the low-pass or smooth part and (ii) the high-pass or detail part, when the length of the signal is N which is divisible by 2. Now suppose $N$ is divisible by $q$ where $1<q \le N$, then we can decompose the signal in $q$ components: one smooth component and $q-1$ detail components. Let $l_{2}(\mathbb{Z}_N)$ be a N-dimensional vector space. Now to generalise the discrete wavelet transform, we define few notations.

\begin{dfn}
For any $z \in {l_2(\mathbb{Z}_{N_1} \times \mathbb{Z}_{N_2} \times \hdots \times \mathbb{Z}_{N_L})}$ ,define \\$\widetilde{z}(n_1,n_2,\hdots,n_L) = \overline{z(N_1-n_1,N_2-n_2, \hdots, N_L-n_L)}$ $\forall$ $n_i$
\end{dfn}

\begin{dfn}
Motivated by the definition given in ~\cite{bib:3}, we define for general case a family of cyclically shifted delay operators as \\$(S_{d_1,d_2,\hdots,d_L}z)(n_1,n_2,\hdots,n_L) = z(((n_1-d_1))_{N_1},((n_2-d_2))_{N_2},\hdots,((n_L-d_L))_{N_L})$\\
This implies that $S_{d_1,d_2,\hdots,d_L}$ shifts $z$ by $d_1,d_2,\hdots,d_L$ circularly to $n_1,n_2,\hdots,n_L$ respectively under modulo operation. This is .
\end{dfn}

\begin{dfn}
Suppose $M\in\mathbb{N}$, $N=qM$ and $u_j \in l_{2}(\mathbb{Z}_N)$ where $j=1,2 \hdots q$. For $n \in \mathbb{Z}$, define system matrix $P(n)$ as

\begin{eqnarray*}
P(n) = \frac{1}{\sqrt q}\left[\begin{array}{cccc}\hat{u_1}(n) & \hat{u_2}(n) & \hdots & \hat{u_q}(n)\\
					    \hat{u_1}(n+M) & \hat{u_2}(n+M) & \hdots & \hat{u_q}(n+M)\\
					   \vdots & \vdots &  & \vdots\\
					      \hat{u_1}(n+(q-2)M) & \hat{u_2}(n+(q-2)M) & \hdots & \hat{u_q}(n+(q-2)M)\\
					    \hat{u_1}(n+(q-1)M) & \hat{u_2}(n+(q-1)M) & \hdots & \hat{u_q}(n+(q-1)M)\end{array}\right]
\end{eqnarray*}
where $\hat{u_j}(n)$ is DFT of $u_j(l)$.
\end{dfn}

We will show that ORPT can be used to generalise DWT to higher dimensions.
Here we will show that for an $L$-dimensional signal we can generate a set of vectors which will span the space and are orthogonal.

\begin{thm}
let $z(n_1,n_2,\dots,n_L)$ be a $L$ dimensional signal.i.e. $z \in l_2(\mathbb{Z}_{N_1} \times \mathbb{Z}_{N_2} \times \hdots \times \mathbb{Z}_{N_L})$ where $N_1 = q_1M_1,N_2 = q_2M_2 \dots N_L=q_LM_L$.

Now define
$z_{k_1}(n_1,n_2,\dots,n_L) = (-1)^{\frac{2 n_1k_1}{q_1}}z(n_1,n_2,\dots,n_L)$\\
$z_{k_2}(n_1,n_2,\dots,n_L) = (-1)^{\frac{2 n_2k_2}{q_2}}z(n_1,n_2,\dots,n_L)$\\
$z_{k_1k_2}(n_1,n_2,\dots,n_L) =  (-1)^{2(\frac{n_1k_1}{q_1}+\frac{n_2k_2}{q_2})}z(n_1,n_2,\dots,n_L)$\\
$\vdots =\vdots$\\
$z_{k_1k_2 \hdots k_L}(n_1,n_2,\dots,n_L) =  (-1)^{2(\frac{n_1k_1}{q_1}+\frac{n_2k_2}{q_2}+ \hdots +\frac{n_Lk_L}{q_L})}z(n_1,n_2,\dots,n_L)$\\\\
where $1 \le k_i \le q_i-1$ and $i = 1,2 \hdots L$
Then the following property holds :

\begin{eqnarray*}
\mathbf{ (i)  } z(n_1,n_2,\dots,n_L)+\sum_{k_1=1}^{q_1-1}z_{k_1}(n_1,n_2,\dots,n_L) +\hdots +\sum_{k_1=1}^{q_1-1}\sum_{k_2=1}^{q_2-1}z_{k_1k_2}(n_1,n_2,\dots,n_L)+ \hdots +\\ \sum_{k_1=1}^{q_1-1}\sum_{k_2=1}^{q_2-1} \hdots \sum_{k_L=1}^{q_L-1}z_{k_1k_2 \hdots k_L}(n_1,n_2,\dots,n_L)\\ =\sum_{k_1=0}^{q_1-1}\sum_{k_2=0}^{q_2-1} \hdots \sum_{k_L=0}^{q_L-1} \exp^{-j2\pi(\frac{n_1k_1}{q_1}+\frac{n_2k_2}{q_2}+ \hdots +\frac{n_Lk_L}{q_L})}z(n_1,n_2,\hdots,n_L)\\
=q_1q_2\hdots q_Lz(n_1,n_2,\hdots,n_L) \text{ when } q_i | n_i
\end{eqnarray*}

\end{thm}

$\mathbf{(ii)} $Let $u_j \in l_2(\mathbb{Z}_{N_1} \times \mathbb{Z}_{N_2} \times \hdots \times \mathbb{Z}_{N_L})$ where $j = 0,1,2,\hdots,(q_1q_2 \hdots q_L-1)$. The following set $B$ forms a orthonormal basis

\begin{eqnarray*}
B = \bigcup_{j=0}^{(q_1q_2 \hdots q_L-1)} \{ \bigcup_{d_1=0}^{(M_1-1)} \bigcup_{d_2=0}^{(M_2-1)}  \hdots \bigcup_{d_L=0}^{(M_L-1)} S_{q_1d_1,q_2d_2,\hdots,q_Ld_L}u_j\}
\end{eqnarray*}

iff $P(n_1,n_2 \hdots , n_L)$ is unitary, where $i^{th}$ column of $P(n_1,n_2 \hdots , n_L)$ can be written as

\begin{eqnarray*}
\frac{1}{\sqrt (q_1q_2\hdots q_L)}\left[\begin{array}{c}\hat{u_i}(n_1,n_2,n_3,\hdots ,n_L)\\
					    \hat{u_i}(n_1+\frac{N_1}{q_1},n_2,n_3,\hdots ,n_L)\\
					   \vdots\\
					   \hat{u_i}(n_1+\frac{(q_1-1)N_1}{q_1},n_2,n_3,\hdots ,n_L)\\
					      \hat{u_i}(n_1,n_2+\frac{N_2}{q_2},n_3,\hdots ,n_L)\\
					      \vdots\\
					      \hat{u_i}(n_1,n_2+\frac{(q_2-1)N_2}{q_2},n_3,\hdots ,n_L)\\
					      \vdots\\
					    \hat{u_i}(n_1+\frac{(q_1-1)N_1}{q_1},n_2+\frac{(q_2-1)N_2}{q_2},n_3+\frac{(q_3-1)N_3}{q_3},\hdots ,\hdots n_L+\frac{(q_L-1)N_L}{q_L})\end{array}\right]
\end{eqnarray*}

or the following conditions are satisfied.

\begin{enumerate}
\item $\sum_{j_1=0}^{q_1-1}\sum_{j_2=0}^{q_2-1} \hdots \sum_{j_L=0}^{q_L-1}\vert {\hat{u_j}(n_1+j_1M_1,n_2+j_2M_2,\hdots,n_L+j_LM_L)}\vert^2 = q_1q_2\hdots q_L$ $\forall$ j
\item $\sum_{j_1=0}^{q_1-1}\sum_{j_2=0}^{q_2-1} \hdots \sum_{j_L=0}^{q_L-1} {\hat{u_j}(n_1+j_1M_1,n_2+j_2M_2,\hdots,n_L+j_LM_L)} \\ \overline{\hat{u_m}(n_1+j_1M_1,n_2+j_2M_2,\hdots,n_L+j_LM_L)} = 0$ for j $\ne$ m 
\end{enumerate}

\begin{prf}
\textbf{Proof of (i) is trivial.}\\
As we know
\begin{eqnarray}
\sum_{k_1=0}^{q_1-1}\sum_{k_2=0}^{q_2-1} \hdots \sum_{k_L=0}^{q_L-1} \exp^{-j2\pi(\frac{n_1k_1}{q_1}+\frac{n_2k_2}{q_2}+ \hdots +\frac{n_Lk_L}{q_L})}(u_l \otimes \widetilde{u_l}) = 
\left \{ \begin{array}{ll}
											 q_1q_2\hdots q_L(u_l \otimes \widetilde{u_l}) &\text{ when } q_i | n_i\\
											 0 & \text{ elsewhere}
											\end{array}\right.
\end{eqnarray}
Since $B$ is a orthonormal set hence we have
\begin{eqnarray}
<u_j,S_{q_1d_1,q_2d_2,\hdots,q_Ld_L}u_j> = \left \{ \begin{array}{ll}
											 1 & \text{ when } d_1=d_2 = \hdots =d_L =0 \\
											 0 & \text{elsewhere}
											\end{array}\right.
\end{eqnarray}
Using equation (5),equation (4) can be written as

\begin{eqnarray}
\sum_{k_1=0}^{q_1-1}\sum_{k_2=0}^{q_2-1} \hdots \sum_{k_L=0}^{q_L-1} \exp^{-j2\pi(\frac{n_1k_1}{q_1}+\frac{n_2k_2}{q_2}+ \hdots +\frac{n_Lk_L}{q_L})}(u_l \otimes \widetilde{u_l}) = q_1q_2\hdots q_L\delta (d_1,d_2,\hdots,d_L)
\end{eqnarray}
Taking DFT of equation $6$ we get,\\\\
$\sum_{j_1=0}^{q_1-1}\sum_{j_2=0}^{q_2-1} \hdots \sum_{j_L=0}^{q_L-1}\vert {\hat{u_j}(n_1+j_1M_1,n_2+j_2M_2,\hdots,n_L+j_LM_L)}\vert^2 = q_1q_2\hdots q_L$ $\forall j$
This proves condition (1).\\
To prove condition (2), from orthonormal set B we have
\begin{eqnarray}
<u_j,S_{q_1d_1,q_2d_2,\hdots,q_Ld_L}u_m> = 0
\end{eqnarray}
We also have 

\begin{eqnarray}
\sum_{k_1=0}^{q_1-1}\sum_{k_2=0}^{q_2-1} \hdots \sum_{k_L=0}^{q_L-1} \exp^{-j2\pi(\frac{n_1k_1}{q_1}+\frac{n_2k_2}{q_2}+ \hdots +\frac{n_Lk_L}{q_L})}(u_j \otimes \widetilde{u_m}) = 
\left\{\begin{array}{ll}
											 q_1q_2\hdots q_L(u_j \otimes \widetilde{u_m}) &\text{ when } q_i | n_i\\
											 0 & \text{ elsewhere}
											\end{array}\right.
\end{eqnarray}
Using equation (8),equation (9) can be written as
\begin{eqnarray}
\sum_{k_1=0}^{q_1-1}\sum_{k_2=0}^{q_2-1} \hdots \sum_{k_L=0}^{q_L-1} \exp^{-j2\pi(\frac{n_1k_1}{q_1}+\frac{n_2k_2}{q_2}+ \hdots +\frac{n_Lk_L}{q_L})}(u_j \otimes \widetilde{u_m}) = 0
\end{eqnarray}
Taking DFT of equation (10) we get,\\\\
$\sum_{j_1=0}^{q_1-1}\sum_{j_2=0}^{q_2-1} \hdots \sum_{j_L=0}^{q_L-1} {\hat{u_j}(n_1+j_1M_1,n_2+j_2M_2,\hdots,n_L+j_LM_L)} \\\overline{\hat{u_m}(n_1+j_1M_1,n_2+j_2M_2,\hdots,n_L+j_LM_L)} = 0$ for j $\ne$ m. This proves condition (2).
Thus $P(n_1,n_2 \hdots , n_L)$ is a unitary matrix.This completes the proof of this theorem.
\end{prf}

A special case of this theorem is given as problem for L = 2 in ~\cite{bib:3},where $q_1=q_2=2$. Now we will show that when $N=qM$ ORPT can be used to define DWT. \\
$B = \bigcup_{j=1}^{q}\{S_{qk}u_j \}_{k=0}^{M-1}$ is an orthonormal basis for $l_{2}(\mathbb{Z}_N)$ then the system matrix $P(n)$ is unitary or the following conditions are satisfied.

\begin{enumerate}
\item $\sum_{i=0}^{q-1} \vert {\hat{u_j}(n+iM)}\vert^2 = q$ for j = 1,2 $\hdots q$
\item $\sum_{i=0}^{q-1}  {\hat{u_j}(n+iM)} \overline{\hat{u_k}(n+iM)} = 0$ for j $\ne$ k 
\end{enumerate}
For example if $x$ is a vector of length $1$x$N$ and $N= 3M$ then we have to choose $R_3$ which is shown in previous section.$u_j$'s can be chosen as columns of $R_3$ after normalisation and appending $N-3$ zeros. $u_j$'s will be of the form \\
\begin{math}
\left(\begin{array}{ccc}\frac{1}{\sqrt{3}} & \frac{2}{\sqrt{6}} & 0 \\\frac{1}{\sqrt{3}} & \frac{-1}{\sqrt{6}} & \frac{1}{\sqrt{2}} \\ \frac{1}{\sqrt{3}} & \frac{-1}{\sqrt{6}} & \frac{1}{\sqrt{2}}\\0 & 0 & 0 \\0 & 0 & 0 \\0 & 0 & 0 \\0 & 0 & 0 \\0 & 0 & 0 \\0 & 0 & 0 \\0 & 0 & 0 \\0 & 0 & 0 \\0 & 0 & 0\end{array}\right)
\end{math}\\
and orthonormal basis $B$ will be in the form of following matrix. Matrix $B$ is applied on $x$ and we get $y=xB$ \\
\begin{math}
\left(\begin{array}{cccccccccccc}\frac{1}{\sqrt{3}}  & 0 & 0 & 0 & \frac{2}{\sqrt{6}} & 0 & 0 & 0 & 0 & 0 & 0 & 0 \\\frac{1}{\sqrt{3}}  & 0 & 0 & 0  & \frac{-1}{\sqrt{6}} & 0 & 0 & 0 & \frac{1}{\sqrt{2}} & 0 & 0 & 0 \\\frac{1}{\sqrt{3}}  & 0 & 0 & 0 & \frac{-1}{\sqrt{6}} & 0 & 0 & 0 & \frac{-1}{\sqrt{2}} & 0 & 0 & 0 \\0 & \frac{1}{\sqrt{3}}  & 0 & 0 & 0 & \frac{2}{\sqrt{6}} & 0 & 0 & 0 & 0 & 0 & 0 \\ 0 &\frac{1}{\sqrt{3}}  & 0 & 0 & 0  & \frac{-1}{\sqrt{6}} & 0 & 0 & 0 & \frac{1}{\sqrt{2}} & 0 & 0  \\0 & \frac{1}{\sqrt{3}}  & 0 & 0 & 0 & \frac{-1}{\sqrt{6}} & 0 & 0 & 0 & \frac{-1}{\sqrt{2}} & 0 & 0 \\0 & 0 & \frac{1}{\sqrt{3}}  & 0 & 0 & 0 & \frac{2}{\sqrt{6}} & 0 & 0 & 0 & 0 & 0 \\0 &  0 &\frac{1}{\sqrt{3}}  & 0 & 0 & 0  & \frac{-1}{\sqrt{6}} & 0 & 0 & 0 & \frac{1}{\sqrt{2}} & 0\\0 & 0 & \frac{1}{\sqrt{3}}  & 0 & 0 & 0 & \frac{-1}{\sqrt{6}} & 0 & 0 & 0 & \frac{-1}{\sqrt{2}} & 0 \\0 & 0 & 0 & \frac{1}{\sqrt{3}}  & 0 & 0 & 0 & \frac{2}{\sqrt{6}} & 0 & 0 & 0 & 0 \\ 0 & 0 &  0 &\frac{1}{\sqrt{3}}  & 0 & 0 & 0  & \frac{-1}{\sqrt{6}} & 0 & 0 & 0 & \frac{1}{\sqrt{2}} \\ 0 & 0 & 0 & \frac{1}{\sqrt{3}}  & 0 & 0 & 0 & \frac{-1}{\sqrt{6}} & 0 & 0 & 0 & \frac{-1}{\sqrt{2}} \end{array}\right)
\end{math}\\
Space formed by first four vectors of $B$ gives the average component of the signal $x$ in $y$. Next four vectors of $B$ gives the first detailed component of the signal whereas the last four vectors gives the second detailed component of the signal.  It can observed that the calculations can be performed by integer operations only without normalising $R_3$. After calculations normalisation factor for each component can be applied i.e. one average and two detailed component.\\
A filter-bank representation of DWT is possible when signal length is divisible by $2$. Likewise a filter-bank representation of a $L$-dimensional signal is also possible,in particular when signal length is divisible by any positive integer $q_i>1,1 \le i \le L$. Defining up-sampler and down-sampler in $L$-dimension.

\begin{dfn}
For any $z \in {l_2(\mathbb{Z}_{N_1} \times \mathbb{Z}_{N_2} \times \hdots \times \mathbb{Z}_{N_L})}$ ,$N_i = q_iM_i,1 \le i \le L$ define $U(z(n_1,n_2,\dots,n_L)) = \left\{\begin{array}{ll}
											 z(\frac{n_1}{q_1},\frac{n_2}{q_2},\hdots,\frac{n_L}{q_L}) &\text{ when } q_i | n_i \forall i \\
											 0 & \text{ elsewhere}
											\end{array}\right\}$\\
											$D(z(n_1,n_2,\dots,n_L)) =  z({n_1}{q_1},{n_2}{q_2},\hdots,{n_L}{q_L})$  \text{$\hspace{1 in}0 \le n_i \le M_i $}
\end{dfn}
\begin{thm}
\textbf{Filterbank representation for L-dimensional case}\\
Suppose $s_i,u_i \in {l_2(\mathbb{Z}_{N_1} \times \mathbb{Z}_{N_2} \times \hdots \times \mathbb{Z}_{N_L})}$, $i = 0,1,2,\hdots,(q_1q_2 \hdots q_L-1)$. For perfect reconstruction
\begin{eqnarray}
\sum_{i=0}^{(q_1q_2 \hdots q_L-1)}\tilde {s_i}*U(D(z*\tilde{u_i})) =z
\end{eqnarray}
for all $z \in {l_2(\mathbb{Z}_{N_1} \times \mathbb{Z}_{N_2} \times \hdots \times \mathbb{Z}_{N_L})}$ if and only if \\
\begin{eqnarray}
P(n_1,n_2 \hdots , n_L)\left[\begin{array}{c} \hat{s_1}\\\hat{s_2}\\ \vdots \\ \hat{s_{(q_1q_2 \hdots q_L-1)}} \end{array}\right]  = \left[\begin{array}{c}\sqrt(q_1q_2 \hdots q_L)\\0\\\vdots\\0\end{array}\right]
\end{eqnarray}
\end{thm}
\begin{prf}
Taking Fourier transform of equation 11 we get,
\begin{eqnarray}
\sum_{i=0}^{(q_1q_2 \hdots q_L-1)}\overline{{\hat{s_i}}} U(D\hat(z*\tilde{u_i})) = \hat z
\end{eqnarray}
Since $U(D(z*\tilde{u_i}))$ can be written as
\begin{eqnarray*}
U(D(z*\tilde{u_i})) = \left\{\begin{array}{ll}
							(z*\tilde{u_i}(l_1,l_2,\hdots,l_L)) & \text{ when } q_i | l_i \forall i \\
							0 & \text{ elsewhere}\end{array}\right\}
\end{eqnarray*}

\begin{eqnarray*}
U(D(z*\tilde{u_i})) = \frac{1}{q_1q_2\hdots q_L}\sum_{k_1=0}^{q_1-1}\sum_{k_2=0}^{q_2-1} \hdots \sum_{k_L=0}^{q_L-1} \exp^{-j2\pi(\frac{l_1k_1}{q_1}+\frac{l_2k_2}{q_2}+ \hdots +\frac{l_Lk_L}{q_L})}z(l_1,l_2,\hdots,l_L)
\end{eqnarray*}
Taking fourier transform of the above equation and substituting in equation 13 ,we get
\begin{eqnarray*}
\frac{1}{q_1q_2\hdots q_L}\sum_{i=0}^{(q_1q_2 \hdots q_L-1)}\overline{{\hat{s_i}}} \sum_{k_1=0}^{q_1-1}\sum_{k_2=0}^{q_2-1} \hdots \sum_{k_L=0}^{q_L-1} \exp^{-j2\pi(\frac{l_1k_1}{q_1}+\frac{l_2k_2}{q_2}+ \hdots +\frac{l_Lk_L}{q_L})}z(l_1,l_2,\hdots,l_L) = \hat z\\
\end{eqnarray*}
or
\begin{eqnarray*}
\frac{1}{q_1q_2\hdots q_L}\sum_{i=0}^{(q_1q_2 \hdots q_L-1)}\overline{{\hat{s_i}}} \sum_{k_1=0}^{q_1-1}\sum_{k_2=0}^{q_2-1} \hdots \sum_{k_L=0}^{q_L-1} \hat{z}(n_1+k_1M_1,n_2+k_2M_2,\hdots,n_L+k_LM_L)\\\overline{\hat{u}}(n_1+k_1M_1,n_2+k_2M_2,\hdots,n_L+k_LM_L) = \hat z
\end{eqnarray*}
This equation leads to equation 12 which proves the theorem.
\end{prf}
It can been observed from equation 12 that a simple perfect reconstruction  can be obtained by putting $s_i = \tilde{u_i},\forall i$. Filter-bank representation of one-dimensional signal is shown below using ORPT.

Similarly filter-bank representation can be given for a signal whose length is divisible by $q^p$. This can be achieved in following way. First DWT based on ORPT is applied on the signal with $R_q$ as described earlier. This will result in one average component and $q-1$ detailed components. Now this is applied on average component again for $p-1$ times.
A $p$-th stage wavelet filter sequence is a sequence of vectors $u_1^1,u_2^1,\hdots,u_q^1,u_1^2,u_2^2,\hdots,u_q^2,\hdots,u_1^p,u_2^p,\hdots,u_q^p$ such that for each $l=1,2,\hdots,p$;$\hspace{0.5in}$$u_i^l \in l_2 (\mathbb{Z}_\frac{N}{q^{l-1}})\forall i$ and the system matrix is of the form
\begin{eqnarray*}
P_l(n) = \frac{1}{\sqrt q}\left[\begin{array}{cccc}{\hat{u}_1^l}(n) & {\hat{u}_2^l}(n) & \hdots & {\hat{u}_q^l}(n)\\
					    {\hat{u}_1^l}(n+\frac{N}{q^l}) & {\hat{u}_2^l}(n+\frac{N}{q^l}) & \hdots & {\hat{u}_q^l}(n+\frac{N}{q^l})\\
					   \vdots & \vdots &  & \vdots\\
					    {\hat{u}_1^l}(n+(q-1)\frac{N}{q^l}) & {\hat{u}_2^l}(n+(q-1)\frac{N}{q^l}) & \hdots & {\hat{u}_q^l}(n+(q-1)\frac{N}{q^l})\end{array}\right]
\end{eqnarray*}
is unitary from theorem $4.1$ for all $n=0,1,2,\hdots,\frac{N}{q^l}-1$ because each stage can be considered independently.\\
For an input $z \in  l_2 (\mathbb{Z}_N)$,define
\begin{eqnarray}
x^1 = D(z*\tilde{u}_1^1) \in (\mathbb{Z}_\frac{N}{q}) \\
y_i^1 = D(z*\tilde{u}_i^1) \in (\mathbb{Z}_\frac{N}{q}) \text{ for } {i =2,3,\hdots,q}
\end{eqnarray}
Here since $z \in  l_2 (\mathbb{Z}_N)$,therefore $U$ and $D$ are taken to be one dimensional upsampler and downsampler.
For $l=2,3,\hdots,p$
\begin{eqnarray}
x^l = D(x^{l-1}*\tilde{u}_1^l) \in (\mathbb{Z}_\frac{N}{q^l}) \\
y_i^l = D(x^{l-1}*\tilde{u}_i^l) \in (\mathbb{Z}_\frac{N}{q^l}) \text{ for } {i =2,3,\hdots,q}
\end{eqnarray}
$x^l$ represents the smoothing information of the signal whereas $y_i^l$ are the detailed information of the signal at the $l$-th stage. This can be further written as
\begin{eqnarray*}
x^l = D(D(\hdots D(D(z*\tilde{u}_1^1)*\tilde{u}_1^2) \hdots *\tilde{u}_1^{l-1})*\tilde{u}_1^l) \\
y_i^l = D(D(\hdots D(D(z*\tilde{u}_1^1)*\tilde{u}_1^2) \hdots *\tilde{u}_1^{l-1})*\tilde{u}_i^l) \text{ for } {i =2,3,\hdots,q}
\end{eqnarray*}
At the output of the $p$-th stage we have one smooth component and $(q-1)$ detailed component. The size of the detailed and smooth components are $\frac{N}{q^{p-1}}$. The output of the $p$-th stage wavelet filter bank is the set of vectors $(y_2^1,y_3^1,\hdots,y_q^1,y_2^2,y_3^2,\hdots,y_q^2,\hdots,y_2^p,y_3^p,\hdots,y_q^p,x_1^p)$.Sum of all the output vectors of the $p$-th stage is 
\begin{eqnarray*}
(q-1)(\frac{N}{q}+\frac{N}{q^2}+\hdots+\frac{N}{q^{p-1}})+\frac{N}{q^{p-1}}=N
\end{eqnarray*}
This is expected in analysis phase. Reconstruction phase (from $l$-th stage to $l-1$ stage)can be described as
\begin{eqnarray*}
\sum_{i=2}^{q}(U(y_i^l))*u_i^l+U(x_l)*u_1^l=x^{l-1}
\end{eqnarray*}
 in similar way to obtain ${l-2}$ from ${l-1}$
\begin{eqnarray*}
\sum_{i=2}^{q}(U(y_i^{l-1}))*u_i^{l-1}+U(x_{l-1})*u_1^{l-1}=x^{l-2}
\end{eqnarray*}
proceeding  in similar way 
\begin{eqnarray*}
\sum_{i=2}^{q}(U(y_i^{1}))*u_i^{1}+U(x_{1})*u_1^{l}=z
\end{eqnarray*}
As we have seen that each stage has its own analysis and synthesis phase. But this process is recursive. Now to do this non recursively we will prove the following useful theorem about interoperability of up-sampler and down-sampler.

\begin{thm}
If $N$ is divisible by $q^l$,let $x,y \in  l_2 (\mathbb{Z}_\frac{N}{q^l})$ and $z \in  l_2 (\mathbb{Z}_N)$.Then

\begin{eqnarray}
D^l(z)*x = D^l(z*U^l(x)) \label{Down:}
\end{eqnarray}

\begin{eqnarray}
U^l(x*y) = U^l(x)*U^l(y)) \label{Up:}
\end{eqnarray}
where
\begin{eqnarray*}
D^l(z) = z(q^ln)\\
U^l(z) = z(\frac{n}{q^l}) = \left \{ \begin{array}{ll}z(n) & \text{ if } {q^l}|{n}\\
									0 & \text{elsewhere}\end{array}\right\}
\end{eqnarray*}
\end{thm}

\begin{prf}
\begin{eqnarray*}
D^l(z)*x = \sum_{m=0}^{\frac{N}{q^l}-1} D^{l}(z)(n-m)x(m)\\
=\sum_{m=0}^{\frac{N}{q^l}-1} z(q^ln-q^lm)x(m)\\
=\sum_{m=0}^{\frac{N}{q^l}-1} z(q^ln-q^lm)U^l(x(q^lm))\\
\end{eqnarray*}
put $k = q^lm$,we get
\begin{eqnarray*}
D^l(z)*x=\sum_{k=0}^{N-1} z(q^ln-k)U^l(x(k))\\
\text{or } D^l(z)*x = D^l(z*U^l(x))
\end{eqnarray*}
This proves equation ~\eqref{Down:}.To prove equation ~\eqref{Up:},as we know
\begin{eqnarray*}
U(x*y) = U(x)*U(y)
\end{eqnarray*}
Applying induction theorem on this we get equation ~\eqref{Up:}.
\end{prf}
Using the above results we will show that the analysis phase of the filter-bank can be computed non-recursively.
\begin{thm}
\textbf{Non recursive computation of recursive filter bank in Analysis phase:} 
Suppose $N$ is divisible by $q^p$. Let $z \in \mathbb{Z}_N$, for $1 \le l \le p,\{u_i^l\}_{i = 1,2,\hdots,q}\in l_2 (\mathbb{Z}_\frac{N}{q^{l-1}})$ define,\\ For l=1 $g^l = u_1^l$,$f_i^l = u_i^l$ for $2 \le i \le q$, \\Then for ${l = 2,3,\hdots,p}$ \\$g^l=g^{l-1}*U^{l-1}(u_1^l)$ and $f_i^l = g^{l-1}*U^{l-1}(u_i^l)$  ${i = 2,3,\hdots,p}$.\\
Prove that  for $l = 1,2,\hdots,p$, $x^l$ and $y_i^l$ are the components of the $l$-th stage recursive filter bank
\begin{eqnarray}
x^l = D^l(z*\tilde{g}^l) \label{analysis_x:}
\end{eqnarray}
\begin{eqnarray}
y_i^l = D^l(z*\tilde{f_i}^l)  \text{ for } {i =2,3,\hdots,q} \label{analysis_y:}
\end{eqnarray}
\begin{prf}
To prove ~\eqref{analysis_x:} and ~\eqref{analysis_y:} applying induction on $l$, for l = 1 as we get first stage smooth and detailed components
\begin{eqnarray*}
x^l = D^l(z*\tilde{g}^l)= D^l(z*\tilde{u}_1^1) \\
y_i^l = D^l(z*\tilde{f_i}^l)= D^l(z*\tilde{u}_i^1)  \text{ for } {i =2,3,\hdots,q}
\end{eqnarray*}
Now assume that equation ~\eqref{analysis_x:} and ~\eqref{analysis_y:} are true for $l-1$ stage. As we know that 
\begin{eqnarray*}
x^l = D(x^{l-1}*\tilde{u}_1^l)
\end{eqnarray*}
Replacing the value of $x^{l-1}$ in this,we get
\begin{eqnarray*}
x^l = D(D^{l-1}(z*\tilde{g}^{l-1})*\tilde{u}_1^l)\\
=D^l(z*\tilde{g}^{l-1}*U(\tilde{u}_1^{l-1})) (\text{using equation } ~\eqref{Down:})\\
=D^l(z*\tilde{g}^{l})
\end{eqnarray*} 
Now for 
\begin{eqnarray*}
y_i^l= D(x^{l-1}*\tilde{u}_i^l) \text{ for } 2\le i \le q\\
=D(D^{l-1}(z*\tilde{g}^{l-1})*\tilde{u}_i^l)\\
=D^l(z*\tilde{g}^{l-1}*U(\tilde{u}_i^{l-1}))\\
=D^l(z*\tilde{f}_i^{l})
\end{eqnarray*}
This proves the theorem.
\end{prf}
\end{thm}
Similarly synthesis phase of the filter-bank can be computed non-recursively.

\begin{thm}
\textbf{Non recursive computation of recursive filter bank in Synthesis phase:} 
In the synthesis phase if the inputs to the $l$-th branch $(1 \le l \le p)$  are $y_i^l$'s ($2 \le i \le q)$, and all other inputs are zero, then the outputs of the synthesis phase are 
\begin{eqnarray}
A_i^l(y_i^l) = f_i^l*U^l(y_i^l) \text{ where } 2 \le i \le q  \label{al:}
\end{eqnarray} 
and if the input to the last branch is $x^p$ and all other inputs are zero then output is
\begin{eqnarray}
B^p(x^p) = g^p*U^p(x^p) \label{bl:}
\end{eqnarray} 
where 
\begin{eqnarray*}
A_i^l(w)= (U(U(\hdots  (U(U(w)*u_i^l)*u_1^{l-1})  \hdots))*u_1^2)*u_1^1 \text{ where } 2 \le i \le q\\
B^l(w) = (U(U(\hdots  (U(U(w)*u_1^l)*u_1^{l-1})  \hdots))*u_1^2)*u_1^1\\
w \in (\mathbb{Z}_\frac{N}{q^{l}})
\end{eqnarray*} 
\begin{prf}
It can be observed that 
\begin{eqnarray*}
B^l(w)= B^{l-1}(U(w)*u_1^l) \\
= U^{l-1}(U(w)*u_1^l)*g^{l-1} \text{ from definition of $g^l$ in Theorem $5.4$ }\\
=U^l(w)*g^l
\end{eqnarray*} 
This proves ~\eqref{bl:} when w = $x^p$ and $l = p$. Similarly
\begin{eqnarray*}
A_i^l(w)= B^{l-1}(U(w)*u_i^l)  \\
= B^{l-2}(U(U(w)*u_i^{l})*u_1^{l-1}) \\
= B^{l-2}(U^2(w)*U(u_i^l)*u_1^{l-1}) \\
=U^{l-2}(U^2(w)*U(u_i^l)*u_1^{l-1}) *g^{l-2}\\
=U^l(w)*U^{l-1}(u_i^l)*U^{l-2}(u_1^{l-1})*g^{l-2} \\
=U^l(w)*U^{l-1}*g^{l-1} \text{ from definition of $g^l and f_i^l$ in Theorem $5.4$ }\\
=U^l(w)*f_i^l
\end{eqnarray*} 
This proves ~\eqref{al:} when w = $y_i^l$.
\end{prf}
\end{thm}
Until now we have shown the non recursive computation of the filter-bank using $f_i^l$'s and $g^l$. Now we will show that the vectors used to find the different components are orthogonal to each other at both the level,intra orthogonality and inter orthogonality. All these vectors together form an orthonormal basis.
\begin{thm}
 The set of vectors $B = (f_2^1,f_3^1,\hdots,f_q^1,f_2^2,f_3^2,\hdots,f_q^2,\hdots,f_2^p,f_3^p,\hdots,f_q^p,g^p)$ forms orthonormal basis.
\end{thm}
\begin{prf}
To prove that $B$ forms an orthonormal basis, in part ($i$) it is proved that $\{\bigcup_{i=2}^{q} \{S_{q^lk}f_i^l\}_{k=0}^{\frac{N}{q^l}-1}\}$ and $\{S_{q^lk}g^l\}_{k=0}^{\frac{N}{q^l}-1}$ are orthonormal for given $l$ (intra level orthonormality). In part ($ii$) it is proved that $\{\bigcup_{i=2}^{q} \{S_{q^lk}f_i^l\}_{k=0}^{\frac{N}{q^l}-1}\}$ and $\{\bigcup_{i=2}^{q} \{S_{q^lk}f_i^m\}_{k=0}^{\frac{N}{q^m}-1}\}$ are also orthonormal(inter level orthonormality).\\
\textbf{part ($i$)}
\begin{eqnarray}
\{\bigcup_{i=2}^{q} \{S_{q^lk}f_i^l\}_{k=0}^{\frac{N}{q^l}-1}\} \cup \{S_{q^lk}g^l\}_{k=0}^{\frac{N}{q^l}-1} \label{e1:}
\end{eqnarray} 
 is orthonormal for $l = 1,2,\hdots,p$. This is proved by induction on $l$. Since 
\begin{eqnarray*}
f_i^1 = u_i^1 (\text{ for } 2 \le i \le q),g^1 = u_1^1
\end{eqnarray*} 
 Now using special case of theorem $5.1 $ for $L=1$ ,
 \begin{eqnarray*}
\{\bigcup_{i=2}^{q} \{S_{qk}f_i^1\}_{k=0}^{\frac{N}{q}-1}\} \cup \{S_{qk}g^1\}_{k=0}^{\frac{N}{q}-1}
\end{eqnarray*} 
is orthonormal. Let's assume that $~\eqref{e1:}$ is true for $l-1$ which means
\begin{eqnarray*}
\{\bigcup_{i=2}^{q} \{S_{q^{l-1}k}f_i^{l-1}\}_{k=0}^{\frac{N}{q^{l-1}}-1}\} \cup \{S_{q^{l-1}k}g^{l-1}\}_{k=0}^{\frac{N}{q^{l-1}}-1}
\end{eqnarray*}
is orthonormal. We know that
\begin{eqnarray}
g^{l-1}*\tilde{g}^{l-1}(q^{l-1}k) = <g^{l-1}, S_{q^{l-1}k} g^{l-1}> = \left\{\begin{array}{ll}1 & \text{if } k=0\\
										       0 & k = 1,2,\hdots,\frac{N}{q^{l-1}}-1\end{array}\right\} \label{e2:}
\end{eqnarray}
Now to prove the  $~\eqref{e1:}$ for $l$
\begin{eqnarray*}
<f_i^{l}, S_{q^{l}k} f_i^{l}> = f_i^{l}*\tilde{f}_i^{l}(q^lk)\\
= g^{l-1}*U^{l-1}(u_i^l)*\tilde{g}^{l-1}*U^{l-1}(\tilde{u}_i^l)(q^lk) \\
= g^{l-1}*\tilde{g}^{l-1}*U^{l-1}(u_i^l*\tilde{u}_i^l)(q^lk) \text{    using ~\eqref{Up:}}
\end{eqnarray*}
or
\begin{eqnarray}
<f_i^{l}, S_{q^{l}k} f_i^{l}> = \sum_{n=0}^{N-1}g^{l-1}*\tilde{g}^{l-1}(q^lk-n)*U^{l-1}(u_i^l*\tilde{u}_i^l)(n)
\end{eqnarray}
As we know
\begin{eqnarray*}
U^{l-1}(u_i^l*\tilde{u}_i^l)(n) = (u_i^l*\tilde{u}_i^l)(m) \text{ when } n =q^{l-1}m \text{ else } 0 \label{e3:}
\end{eqnarray*}
substituting this back in $~\eqref{e2:}$, we get

\begin{eqnarray}
<f_i^{l}, S_{q^{l}k} f_i^{l}> = \sum_{m=0}^{\frac{N}{q^{l-1}}-1}g^{l-1}*\tilde{g}^{l-1}(q^lk-q^{l-1}m)*(u_i^l*\tilde{u}_i^l)(m)
\end{eqnarray}

Using $~\eqref{e2:}$,we get
\begin{eqnarray*}
<f_i^{l}, S_{q^{l}k} f_i^{l}> = (u_i^l*\tilde{u}_i^l)(qk)=\left\{\begin{array}{ll}1 & \text{if } k=0\\
									 0 & k = 1,2,\hdots,\frac{N}{q^{l}}-1\end{array}\right\}
\end{eqnarray*}
From this it follows that 
\begin{eqnarray}
\{\bigcup_{i=2}^{q} \{S_{q^lk}f_i^l\}_{k=0}^{\frac{N}{q^l}-1}\} \label{e4:}
\end{eqnarray} 
is orthonormal.Proceeding in similar way,we get
\begin{eqnarray}
<g^{l}, S_{q^{l}k} g^{l}> = (u_1^l*\tilde{u}_1^l)(qk)=\left\{\begin{array}{ll}1 & \text{if } k=0\\
									 0 & k = 1,2,\hdots,\frac{N}{q^{l}}-1\end{array}\right\} \label{e5:}
\end{eqnarray}
and
\begin{eqnarray}
<f_i^{l}, S_{q^{l}k} g^{l}> =0 \label{e6:}
\end{eqnarray}
Combining $~\eqref{e4:}$,$~\eqref{e5:}$ and $~\eqref{e6:}$ proves $\eqref{e1:}$ for $l=1,2,\hdots,p$. \\
\textbf{part ($ii$)}
Now to prove that the subsets at any level are also orthonormal i.e. $\{\bigcup_{i=2}^{q} \{S_{q^lk}f_i^l\}_{k=0}^{\frac{N}{q^l}-1}\}$ and $\{\bigcup_{i=2}^{q} \{S_{q^mk}f_i^m\}_{k=0}^{\frac{N}{q^m}-1}\}$ are orthonormal($m\ne l$). Let us assume that
\begin{eqnarray}
V^l = span\{S_{q^lk}g^l\}_{k=0}^{\frac{N}{q^l}-1} \\
W_i^l = span\{S_{q^lk}f_i^l\}_{k=0}^{\frac{N}{q^l}-1} \text{ for } 2 \le i \le q
\end{eqnarray}
from previous section we know that 
\begin{eqnarray*}
V^l \bot W_1^l \bot W_2^l \hdots \bot W_q^l  
\end{eqnarray*}
is true. This means that subspaces are orthonormal at any given $l$.
Claim
\begin{eqnarray}
V^l \oplus W_1^l \oplus W_2^l \hdots \oplus W_q^l   = V^{l-1} \label{e7:}
\end{eqnarray}
To prove this we have to show that  $V^l$ and $W_i^l$ are subspaces of $V^{l-1}$. We know that
\begin{eqnarray*}
S_{q^lk}g^l(n) = g^l(n-q^lk) \\
= g^{l-1}*U^{l-1}(u_1^l)(n-q^lk)\\
=\sum_{m=0}^{N-1}g^{l-1}(n-q^lk-m)U^{l-1}(u_1^l)(m)\\
=\sum_{r=0}^{\frac{N}{q^{l-1}}-1}g^{l-1}(n-q^lk-q^{l-1}r)u_1^l(r)\\
\end{eqnarray*}
\begin{eqnarray}
S_{q^lk}g^l(n) =\sum_{r=0}^{\frac{N}{q^{l-1}}-1}u_1^l(r)S_{q^{l-1}(r+qk)}g^{l-1}(n) \label{e8:}
\end{eqnarray}
Similarly we get
\begin{eqnarray}
S_{q^lk}f_i^l(n) =\sum_{r=0}^{\frac{N}{q^{l-1}}-1}u_i^l(r)S_{q^{l-1}(r+qk)}g^{l-1}(n) \text{ for } 2 \le i \le q \label{e9:}
\end{eqnarray}
Combining $~\eqref{e8:}$ and $~\eqref{e9:}$ we observe that $V^l$ and $W_i^l$ are subspaces of $V^{l-1}$. Now
\begin{eqnarray*}
dim(V^l)= \frac{N}{q^l}\\
dim(W_i^l)= \frac{N}{q^l} \text{ for } 2 \le i \le q\\
dim(V^l) + \sum_{i=2}^{q}dim(W_i^l) = \frac{N}{q^{l-1}}
\end{eqnarray*}
which is the dimension of $V^{l-1}$. Hence claim is proved. $~\eqref{e7:} $ can be also be written as
\begin{eqnarray*}
 V^{l-1} = V^l \oplus W_1^l \oplus W_2^l \hdots \oplus W_q^l  \\
 = V^{l+1}\oplus\sum_{i=2}^{q}W_i^{l+1}\oplus\sum_{i=2}^{q}W_i^{l}
\end{eqnarray*}
Since $W_i^l$ and $V^{l-1}$ are orthogonal, $V^l$ and $V^{l-1}$ are also orthogonal, hence this proves the theorem that the set $B$ is orthonormal. Therefore
\begin{eqnarray*}
 V^{0} = V^1 \oplus \sum_{i=2}^{q}W_i^{1}  \\
 V^{1}= V^{2} \oplus\sum_{i=2}^{q}W_i^{2}
\end{eqnarray*}
Therefore $V^0$ can be written as
\begin{eqnarray*}
 V^{0} = V^{2} \oplus\sum_{i=2}^{q}W_i^{1} \oplus \sum_{i=2}^{q}W_i^{2}  \\
\end{eqnarray*}
In general,
\begin{eqnarray*}
 V^{0} = V^{p-1} \oplus\sum_{i=2}^{q}W_i^{1} \oplus \sum_{i=2}^{q}W_i^{2} \hdots   \oplus\sum_{i=2}^{q}W_i^{p-1}
\end{eqnarray*}
\end{prf}
It can be observed that DWT based on ORPT also satisfies nesting property i.e.  $V^{j} \subset V^{j-1}$. It can also be observed that this also satisfies density, separation and scaling property of DWT in finite dimensional discrete signals. It is clear that Haar wavelet is a special case of ORPT  where $R_2$ is being used.DWT using ORPT can be extended to continuous domain also.
\section{Results}
In this section we will show the application of ORPT based DWT on different images. Now suppose we have an image $x$ of size $N$x$N$, where $N=qM$.\\ Matrix $B$ is constructed based on $R_q$ as described in the example. Now this $B$ can be applied to the columns and rows. Figure $1$ shows the input image. Application  of $R_3$  on this image will result in figure $2$. Cumulative energy plot with application of different operation is shown in figure $3$.

\begin{figure*}
\begin{center}
\includegraphics[height = 3.2in]{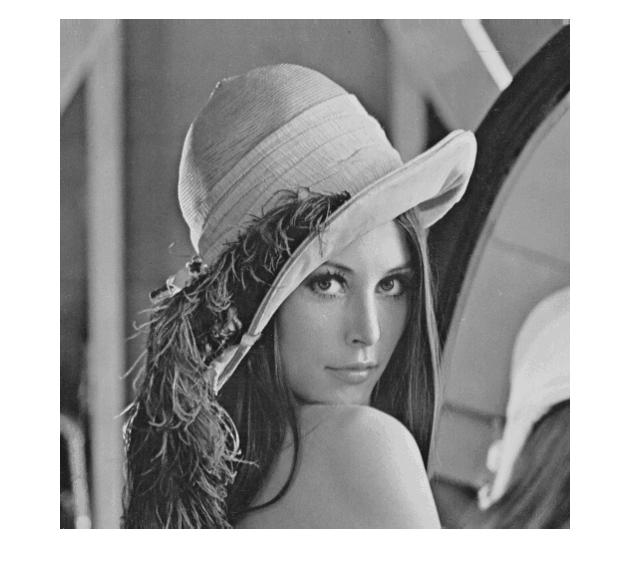}
\caption{Original Image}
\end{center}
\end{figure*}

\begin{figure*}
\begin{center}
\includegraphics[height = 4.2in]{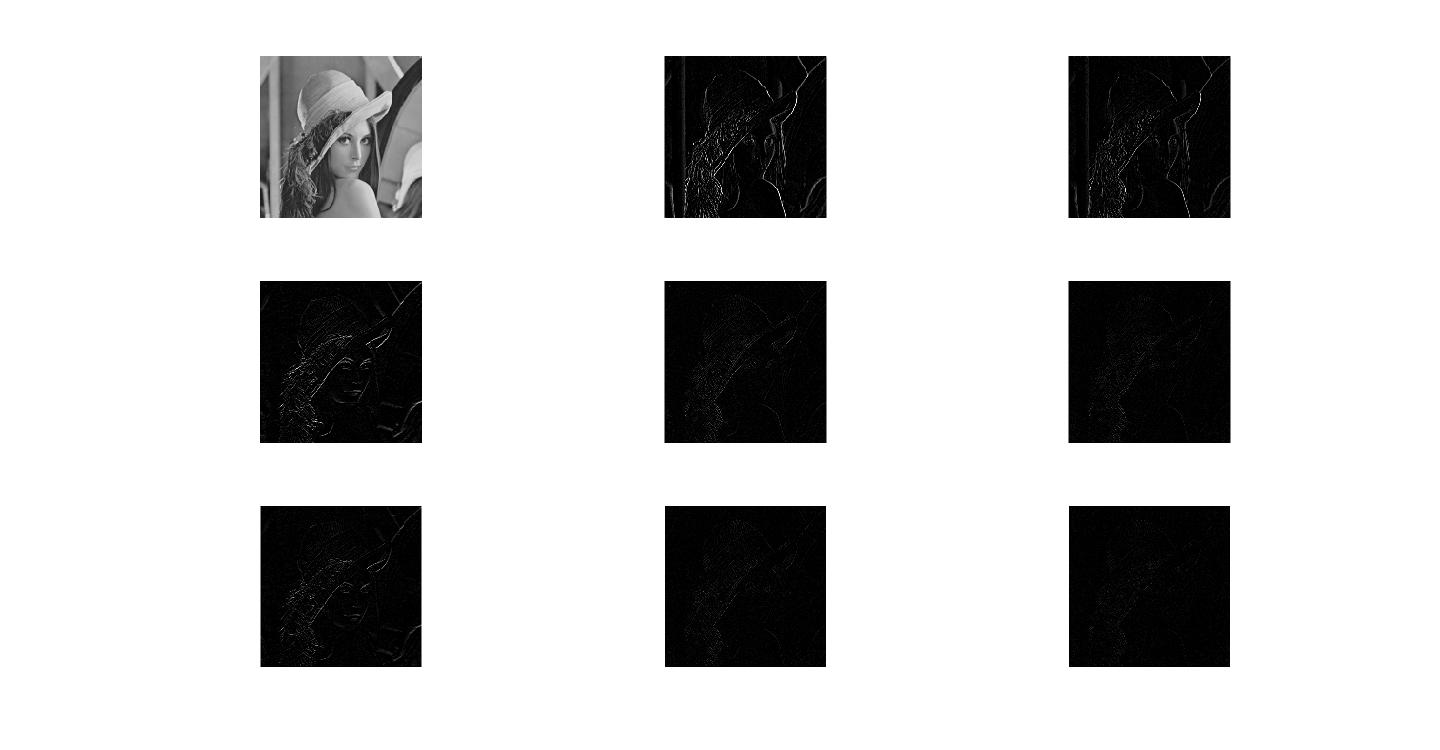}
\caption{Application of ORPT matrix to the rows and columns of a  two dimensional image}
\end{center}
\end{figure*}

\begin{figure*}
\begin{center}
\includegraphics[height = 3.5in]{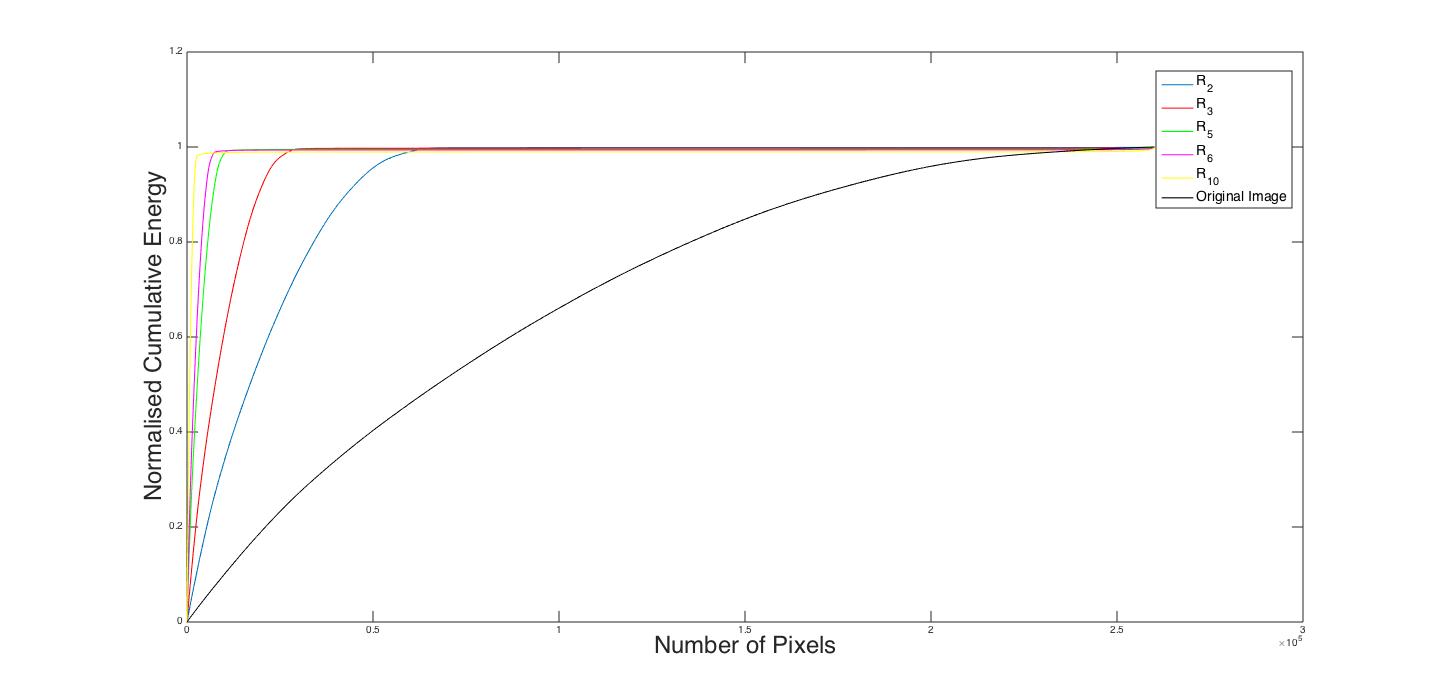}
\caption{Cumulative energy of original image(black) after applying $R_2$(blue),$R_3$(red),$R_5$(green),$R_6$(yellow),$R_{10}$(magenta)}
\end{center}
\end{figure*}
%


\section{Concluding Remarks}
In this paper we have defined Orthogonal Ramanujan Sums which are based on Ramanujan Sums. Some of its properties are discussed in this paper. Orthogonal Ramanujan Periodicity Transform is defined based on ORS. Another application of ORS is demonstrated in MRA, where it is shown that this can be used to generate MRA at any scale.

\end{document}